\documentclass[twocolumn, floatfix, superscriptaddress]{revtex4-1}
\usepackage{amsmath, amssymb}
\usepackage{txfonts}
\usepackage{graphicx} 
\usepackage{hyperref}

\newcommand{\ham}{\mathcal{H}}
\newcommand{\expct}[1]{\langle #1 \rangle}
\newcommand{\cket}[1]{|#1\rangle}
\newcommand{\bra}[1]{\langle #1|}
\begin{document}
\title{Origin of Biquadratic Exchange Interactions in a Mott Insulator as a Driving Force of Spin Nematic Order}
\author{Katsuhiro Tanaka}
\affiliation{Department of Basic Science, University of Tokyo, 3-8-1 Komaba, Meguro, Tokyo 153-8902, Japan}
\author{Yuto Yokoyama}
\affiliation{Department of Basic Science, University of Tokyo, 3-8-1 Komaba, Meguro, Tokyo 153-8902, Japan}
\author{Chisa Hotta}
\affiliation{Department of Basic Science, University of Tokyo, 3-8-1 Komaba, Meguro, Tokyo 153-8902, Japan}
\begin{abstract}
We consider a series of Mott insulators in unit of two orbitals each hosting spin-1/2 electron, 
and by pairing two spin-1/2 into spin-1 triplet, derive the effective exchange interaction between the 
adjacent units via fourth order perturbation theory. 
It turns out that the biquadratic exchange interaction between spin-1, which is one of the origins of the nematic order, 
arises only in processes where the four different electrons exchange cyclically along the twisted loop, 
which we call ``twisted ring exchange'' processes. 
We show that the term becomes the same order with the Heisenberg exchange interactions when the on-orbital Coulomb 
interaction is not too large. 
Whereas, the inter-orbital Coulomb interactions give rise to additional processes that cancel the twisted ring exchange, and strongly suppresses the biquadratic term. 
The Mott insulator with two electrons on degenerate two orbitals is thus not an ideal platform to study such nematic orders. 
\end{abstract}
\maketitle
Spin nematics is a phase of matter without magnetic order, 
but still breaks the spin-rotation symmetry~\cite{nmtc_biqua_01,nmtc_biqua_03}. 
This exotic state established itself as an intermediate category of magnetism and quantum spin liquids, 
with an advantage in that it could be captured much more easily than the spin liquids by a signature of symmetry breaking, 
both in material systems and in toy lattice models. 
A more general description of spin nematics is ``a quadrupole order of quantum $S=1$'', 
and thus, the order parameter is a symmetric and traceless rank-2 tensor operator given by 
$\hat{Q}^{\alpha\beta}_j=\hat{S}^\alpha_j \hat{S}^\beta_j+ \hat{S}^\beta_j \hat{S}^\alpha_j -2S_j(S_j+1)/3 \delta_{\alpha\beta}$, 
where we denote $S^\alpha_j$ ($\alpha=x,y,z$) as $\alpha$-component of $j$-th spin-1, 
and discriminate it from $s^\alpha_j$ which is the spin-1/2 operator we see later. 
A basic Hamiltonian that naturally realizes this order is a bilinear-biquadratic Hamiltonian of $S=1$; 
\begin{align}
\ham_{\rm BB}= \sum_{\langle i,j\rangle} \left[ J \hat S_i \cdot \hat S_j + B (\hat S_i\cdot \hat S_j)^2 \right],
\label{hambb}
\end{align}
where the sum $\langle i,j\rangle$ runs over nearest-neighbor pairs of spin-1. 
Since $\hat Q_i \cdot \hat Q_j=2(\hat S_i\cdot \hat S_j)^2 + \hat{S}_i\cdot \hat S_j - 2S^2(S+1)^2/3$ \cite{nmtc_ring_03}, 
the above Hamiltonian is transformed to 
$\ham_{\rm BB} = \sum_{\langle i,j\rangle} \left[ (J-B/2) \hat S_i \cdot \hat S_j + B/2 \hat Q_i \cdot \hat Q_j \ (+\text{const.}) \right]$, 
and the relative strength of coupling constants of the two competing terms, $B/J$, may determine 
the ground state to be either magnetic or nematic. 
In a one dimensional chain, a rich phase diagram~\cite{chubukov} predicted a possible nematic order between the ferro and antiferromagnets, 
but it turned out to be transformed to a dimerized phase, a sort of valence bond solid of nematic order, 
due to large quantum fluctuation~\cite{moessner,grover,harada07,hu}. 
In two dimensional square and triangular lattices, the ferro-quadrupolar phase is found at $B/J \lesssim -1$, 
and also a stable antiferro-quadrupolar ones in the latter lattice at $B/J \gtrsim 1$~\cite{tsunetsugu,nmtc_biqua_05,nmtc_biqua_02}. 
\par
Unfortunately, however, the value of $|B/J|$ required for the nematic phases to appear in an $S=1$ spin system 
is larger than one, which is seemingly rather too large for a simple Mott insulator to realize. 
However, the spin-1/2 Heisenberg model with competing nearest ferromagnetic and next nearest antiferromagnetic exchanges 
yields a small nematic phase as a two magnon bound state~\cite{nmtc_j1j2_00}. 
The nematic phase also appears near the saturation field in a spin-1/2 ladder with diagonal Heisenberg and ring exchange interactions~\cite{hikihara}. 
Thus, decomposing the spin-1 into a pair of spin-1/2 may give a good reason for $|B|$ to become large. 
We show that the twisted ring exchange interaction originating from the fourth order perturbation terms in the Mott insulator indeed fits this scenario, 
and gives rise to the effective biquadratic interaction of the same order with the antiferromagnetic bilinear interaction. 
%
\begin{figure}
\centering
\includegraphics[width=8.5cm]{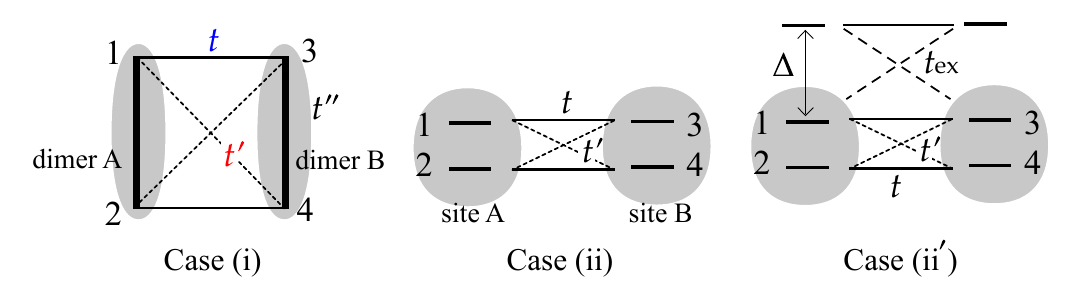}
\caption{Mott insulators hosting one electron per orbital, where the orbital-1 and 2, orbital-3 and 4 form pair-A and B, respectively. 
Case (i) considers each pair of orbitals as ``dimer'' and Case (ii) as ion (site) having degenerate orbitals which is realized in the 
$d$-electron systems. 
Case (ii') discussed in Ref.~\cite{4ptbtn_02} takes account of an additional quasi-degenerate energy level, 
where the loop similar to process-T between the excited states yields a biquadratic term of the same order with a bilinear term. 
}
\label{f1}
\end{figure}
\par
As a starting point, we consider a unit of two orbitals each occupied by a single electron, 
and prepare two sets of such orbitals. 
This situation could be realized in two different cases as shown in Fig.~\ref{f1}; 
(i) four different sites each with a single orbital, and 
(ii) two different sites each having degenerate two orbitals. 
We consider a strong coupling parameter region, 
namely when the electron hopping between orbitals is much smaller than the electronic interactions. 
We first perform a perturbation up to fourth order using Schrieffer--Wolff transformation~\cite{schrieffer-wolff} which keeps the unitarity of the resultant effective Hamiltonian, and derive an effective interaction between spin-1/2's per orbital. 
Then, by pairing the spin-1/2, we project these effective interactions to 
the restricted Hilbert space consisting only of triplets (spin-1), and obtain the form, Eq.~(\ref{hambb}). 
This treatment extracts the magnetic interaction between spin-1's on neighboring sites, 
while notice that it applies not only to the spin-1 lattice models, 
but to models whose low energy states host both triplets and singlets. 
Whenever the triplets become neighbors, they interact magnetically as Eq.~(\ref{hambb}) with evaluated values of $B$ and $J$. 
\par
Let us first deal with the simplest Case (i), starting from the single-band Hubbard Hamiltonian with half-filling; 
\begin{align}
\ham_0= - \sum_{\expct{i, j}, \sigma} t_{ij} \left(c_{i\sigma}^{\dag} c_{j\sigma} + \text{h.c.}\right)
        + \sum_i U n_{i\uparrow} n_{i\downarrow}, 
\label{ham0}
\end{align}
where $c_{i\sigma}^\dag$/$c_{i\sigma}$ is the creation/annihilation operator on site-$i$ with spin $\sigma$, 
and $n_{i\sigma}=c_{i\sigma}^{\dag} c_{i\sigma}$ is the number operator. 
We consider three different species of transfer integrals, $t$, $t'$ and $t''$ shown in Fig.~\ref{f1}. 
When $U$ is strong enough, each orbital is occupied by a single electron, and the low energy degrees of 
freedom is the spin-1/2 on each orbital, $s_i^z=\pm 1/2$. 
The effective Hamiltonian at second order perturbation at $t_{ij}/U \ll 1$ is the well-known Heisenberg term, 
$\ham^{(2)}_{\rm eff}=\sum_{\langle i,j\rangle} J_{ij}s_i\cdot s_j$, with $J_{ij}=4t_{ij}^2/U$. 
The third order terms cancel out on the whole. 
\par
\begin{figure}
\centering
\includegraphics[width=8cm]{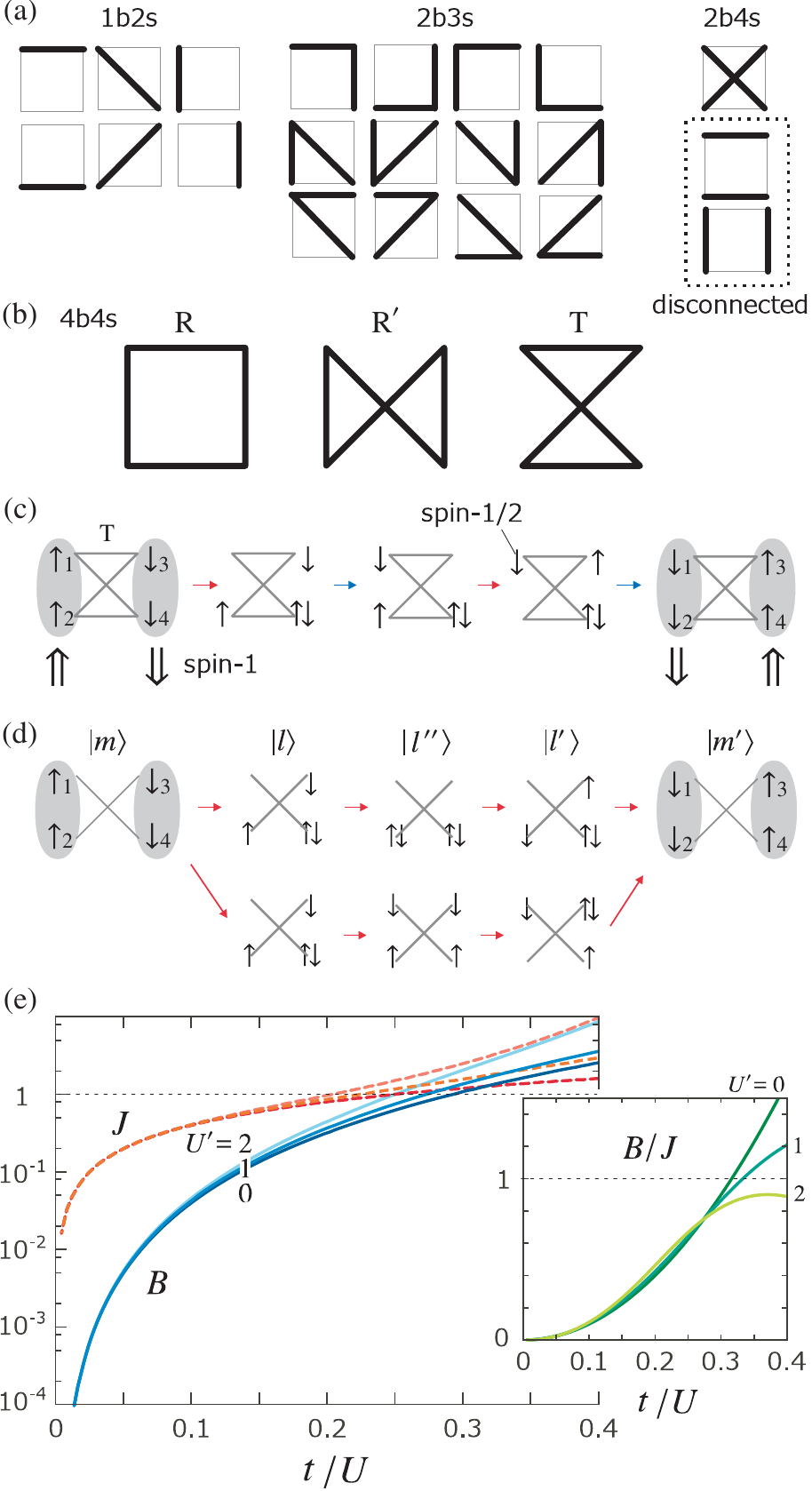}
\caption{(a) Fourth order processes over less than three bonds. 
(b) Three different (4b4s) processes consisting of closed loops, which refer to the typical ring exchanges (R, R') 
and twisted ring exchange (T). 
The processes require all the four hopping terms along these loops to be present. 
(c) One of the processes that contribute to the spin-1 biquadratic exchange, where $\cket{\Uparrow_A\Downarrow_B}$ flips to 
 $\cket{\Downarrow_A\Uparrow_B}$. 
(d) Typical disconnected processes that cancel out and do not contribute to $B$ at $U'=0$. 
(e) Evaluation of model parameters, $J$ and $B$, of Eq. (\ref{jb_all}), in Case (i) based on spin-1 degrees of freedom per site, 
with $U'=0,1,2$ and $t=t'=t''=1$ as a function of $t/U$. 
The contributions at $U'\ne 0$ in Eqs.~(\ref{BV})--(\ref{J2b3sU'}) are included. 
}
\label{f2}
\end{figure}

The fourth order terms are classified into four categories shown in Figs.~\ref{f2}(a) and \ref{f2}(b) as 
$\ham^{(4)}_{\rm eff}\!=\!\ham^{\rm (1b2s)}_{\rm eff}\!+\!\ham^{\rm (2b3s)}_{\rm eff}\!+\!\ham^{\rm (2b4s)}_{\rm eff}\!+\!\ham^{\rm (4b4s)}_{\rm eff}$; 
hopping processes taking place along one, two, and four different bonds (1b, 2b, 4b) and 
over two to four different sites (2s, 3s, 4s), are denoted as (1b2s), (2b3s), (2b4s), and (4b4s), respectively. 
The (1b2s) and (2b3s) contribute to the Heisenberg interaction, and all the (2b4s) cancel out. 
The (4b4s) consisting of hoppings along four all different bonds 
yields the so-called ring exchange term, 
\begin{align}
&\ham^{\rm (4b4s)}_{\rm eff}
=-\frac{4K_{\cal C}}{5}\sum_{(i<j) \in a,b,c,d}s_{i}\cdot s_{j} \nonumber\\
& +4K_{\cal C}\!\!\!\sum_{[a-b-c-d]}\!\!\!\!\!
\Bigl[(s_{a}\cdot s_{b})(s_{c}\cdot s_{d})+(s_{a}\cdot s_{d})(s_{b}\cdot s_{c})-(s_{a}\cdot s_{c})(s_{b}\cdot s_{d})\Bigr],\nonumber\\
&\hspace{5mm} K_{\cal C}=20t_{ab}t_{bc}t_{cd}t_{da}/U^{3}, 
\label{h4ringex}
\end{align}
where we take the hoppings along the closed loop $\mathcal{C}$, consisting of $a\!-\!b\!-\!c\!-\!d\!-\!a$. 
As shown in Fig.~\ref{f2}(b), there are three different closed paths of fourth order that contribute to 
$\ham^{\rm (4b4s)}_{\rm eff}$; 
the first one using $t$ and $t''$ along $(a,b,c,d)=(1,2,4,3)$ (${\cal C}=$ R), which we call process-R, 
is a typical ring exchange first discussed by Takahashi~\cite{4ptbtn_01}. 
The one along $(1,2,3,4)$ is denoted as process-R', and the last one $(1,3,2,4)$ as process-T. 
The contributions to the effective Hamiltonian from these processes are given by assigning the indices of spins $a-d$ in Eq.~(\ref{h4ringex}) 
the orbital indices $1-4$ along the closed paths of ${\cal C}=$ R, R' and T. 
The processes-R' and T are derived already by Calzado and Malrieu as extra four-body-spin exchange processes~\cite{4ptbtn_03}. 
Equation~(\ref{h4ringex}) differs by a factor in the first term from the ring exchange in solid $^3$He~\cite{he3_01,ceperley}, 
which is described by $(P_4 + P^{-1}_4)$ with an operator $P_4$ permutating the spins clockwise. 
\par
The next step is to transform the above expression of $\ham^{(n)}_{\rm eff}$ ($n=2,4$) 
by the spin-1/2 degrees of freedom defined on each orbital 
into the spin-1 degrees of freedom on each pair of orbitals. 
This is done by projecting the effective Hamiltonian to a basis of triplets using the projection operator, ${\cal P}_1$, as 
$\tilde \ham^{(n)}_{\rm eff}={\cal P}_1 \ham^{(n)}_{\rm eff} {\cal P}_1$. 
We then find an effective Hamiltonian of the form of Eq.~(\ref{hambb}) with the coupling constants given separately for each process as 
\begin{align}
& J = J_2+\sum_{\rm process}J_4^{\:({\rm process})}, \hspace{3mm} B=\sum_{\rm process}B_4^{\:({\rm process})}, 
\label{jb_all}
\end{align}
where the subscript indicates the order of perturbation, and the ``process'' indicates (1b2s), (2b3s), etc. 
For the Heisenberg terms, we find 
\begin{align}
& J_2= 2 (t^2+t'^2)/U, 
\label{J2}\\
& J_4^{\rm (1b2s+2b3s)}= (-8t^4-8t'^4+4t^2t''^2+4t'^2t''^2)/U^3, 
\label{J2b3s}
\\
& J_4^{\rm (R)}=-K_{\rm R}/5 =-4 t^2t''^2 / U^{3}, 
\label{JR}
\\
& J_4^{\rm (R')}=-K_{\rm R'}/5=-4 t'^2t''^2/U^{3}, 
\label{JR2}
\\
& J_4^{\rm (T)}= 4K_{\rm T}/5= 16t^2t'^2/U^3. 
\label{JT}
\end{align}
At this stage, we see that the contributions from (4b4s) severely depend on the geometry of paths, 
and in fact, the biquadratic term of spin-1 appears only in process-T as 
\begin{align}
B_4^{\rm (T)}&= 2K_{\rm T} = 40t^2t'^2/U^3. 
\label{BT}
\end{align}
This could be understood more intuitively as follows; 
let us explicitly show a matrix representation of the biquadratic term in Eq.~(\ref{hambb}) between spin-A and spin-B 
within the $S^z_A+S^z_B=0$ space as 
\begin{equation}
\bra{m}\:(\hat S_A\cdot \hat S_B)^2 \cket{m'}=
\begin{pmatrix}
2 & -1 & 1 \\
-1 & 2  & -1 \\
1 & -1 & 2 \\
\end{pmatrix},
\label{mat}
\end{equation}
where the three basis states are chosen as $\cket{m}=\cket{\!\Uparrow_A\Downarrow_B}$,  
$\cket{0_A0_B},\cket{\Downarrow_A\Uparrow_B}$, 
with $\Uparrow, 0,\Downarrow$ indicating $S^z=1,0,-1$ of each spin-1, respectively. 
One of the main roles of the spin-1 biquadratic exchange term is to flip the pairs as, 
$\cket{\!\Uparrow_A\Downarrow_B}$ to $\cket{\Downarrow_A\Uparrow_B}$, and vice versa ((1, 3) and (3, 1) elements of Eq.~(\ref{mat})).
Here, decomposing these $S=1$ and $S^z=\pm 1$ spins into the triplets of $s^z=\pm 1/2$ on orbitals 1--4, as 
$\cket{\Uparrow_A}=\cket{\uparrow_1 \uparrow_2}$ and $\cket{\Downarrow_B}=\cket{\downarrow_3\downarrow_4}$, 
we recall the perturbation processes. 
As shown in Fig.~\ref{f2}(c), part of the process-T contributes to this spin flip. 
If we try to do the same thing in process-R and R', it immediately breaks down. 
This is because in order to make this flip, we need to transfer the up spin electrons on orbital-1 and 2 to orbital-3 and 4, 
and the down spin electrons on orbital-3 and 4 to orbital-1 and 2, while to do so via ring exchange process, 
orbital-1 needs to be connected with both orbital-3 and 4 and so as orbital-2 along the loop. 
One may think that the (2b4s) processes, hopping back and forth along the two different bonds, 
may also flip the spins in the above mentioned manner. 
However, such processes are basically a combination of two independent second order exchange processes, 
which we call ``disconnected processes'', and cancel out on the whole (see Fig.~\ref{f2}(d)). 
This is natural because, otherwise, arbitrary choices of two independent bonds in the bulk system will generate numbers of magnetic long range interactions 
no matter how distant they were separated. 
Thus, we finally find that the twisted ring exchange (T) is responsible for the biquadratic interaction, 
whereas the ordinary ring exchanges (R and R') do not. 
This is the main message to deliver in the present paper.
As shown in Fig. \ref{f2}(e), the evaluated $B$ and $J$ in  Eq.~(\ref{jb_all})~\cite{supplement1}, take the same order 
when $U\lesssim 5t$, which is not too unrealistic. 
\par
Notice that, although we projected out the $S=0$ (singlet) state of each site, there are finite terms between singlets and triplets. 
Namely, ${\cal P}_1 \ham^{(n)}_{\rm eff} (1-{\cal P}_1)$ and $(1-{\cal P}_1)\ham^{(n)}_{\rm eff}(1-{\cal P}_1)$ are not at all negligible 
both at $n=2$ and 4. 
However, these terms work to control the population of triplets and singlets, which will be discussed elsewhere~\cite{yyoko}, 
and thus only indirectly contribute to the magnetic properties as they do not yield any magnetic exchange interaction. 
\par
One extension of Case (i) is to add the intra-dimer Coulomb interactions, $\ham_I= U'(n_1n_2+n_3n_4)$, to Eq.~(\ref{ham0}). 
After performing the same perturbation calculation, one finds that Eqs.~(\ref{J2})--(\ref{JR2}) do not change. 
However, there emerges an additional contribution from the disconnected (2b4s) processes as 
\begin{equation}
B_4^{({\rm 2b4s};U')}\!=\!\frac{4(t^4+t'^4)}{U^2}\left(\frac{2}{U}-\frac{1}{U-U'}-\frac{1}{U+U'}\right). 
\label{BV}
\end{equation}
This is simply because, at finite $U'$, the two bonds are no longer disconnected. 
The ones from process-T are corrected from Eqs.~(\ref{JT}) and (\ref{BT}), and $J_4^{{\rm (2b3s;}U'{\rm)}}$ is added as 
\begin{align}
& J_4^{\rm (T)} = 16t^2 t'^2/U^2(U-U'), 
\label{JTU'}
\\
& B_4^{\rm (T)} = t^2 t'^2\left[32/U^2(U-U')+8/U^2(U+U')\right], 
\label{BTU'}
\\
& J_4^{{\rm (2b3s;}U'{\rm)}}= 16t^2t'^2\left[-1/U^3+1/U^2(U-U')\right]. 
\label{J2b3sU'}
\end{align}
\par
By further adding the Hund interaction between the dimerized two orbitals, we find Case (ii). 
We consider a Kanamori Hamiltonian\cite{kanamori,georges} $\ham=\ham_0+ \ham_d$, with 
\begin{align}
\ham_d&= \sum_{j\ni (a \ne b)}\left( U' - J_{H} \right) \sum_{\sigma} n_{a\sigma} n_{b\sigma}
	+ U' \sum_{i}\left(n_{a\uparrow} n_{b\downarrow} + n_{a\downarrow} n_{b\uparrow}\right) \nonumber \\
& \;\;+ J_{H}\left(c_{a\uparrow}^{\dag} c_{b\uparrow} c_{b\downarrow}^{\dag} c_{a\downarrow} 
	+ c_{a\downarrow}^{\dag} c_{b\downarrow} c_{b\uparrow}^{\dag} c_{a\uparrow} \right)
\nonumber \\
& \;\;+ J_{p}\left(c_{a\uparrow}^{\dag} c_{b\uparrow} c_{a\downarrow}^{\dag} c_{b\downarrow}
+ c_{b\downarrow}^{\dag} c_{a\downarrow} c_{b\uparrow}^{\dag} c_{a\uparrow}
	\right), 
\end{align}
where the orbital indices $(a,b) = (1,2), (3,4)$ are those on the same site. 
The inter-orbital intra-site Coulomb interaction $U'$ and the Hund coupling $J_H$ are taken as such that
they fulfill $U=U'+2J_H$, in crystal fields of cubic symmetry. 
Therefore, Case (ii) roughly corresponds to the large-$U'$ version of Case (i). 
The $J_p$-term expresses the pair hopping. 
As the two degenerate pairs of orbitals are orthogonal, 
we set $t''=0$, which has the geometry of the twisted ring exchange. 
\begin{figure}
	\centering
	\includegraphics[width=8.5cm]{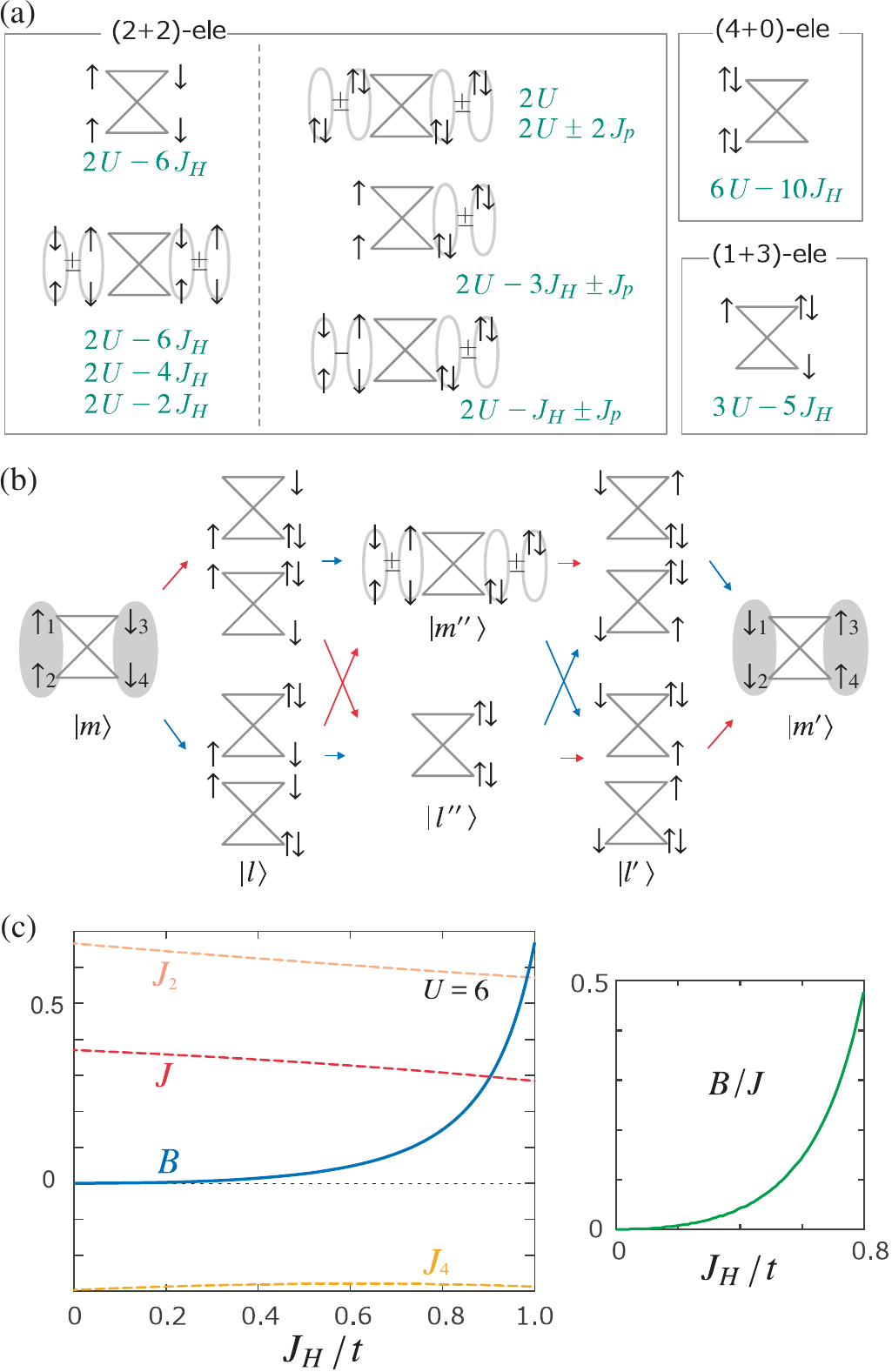}
	\caption{(a) Classification of states; (2+2)-ele (low energy states $\cket{m}$) and 
		(1+3), (4+0)-ele (excited states $\cket{l}$), and the list of their energy values. 
		There are many other configurations not shown, having the same energies as listed. 
		The right half of the (2+2)-ele states have the double occupancy of orbitals. 
		(b) Twisted ring exchange processes modified from those of Fig.~\ref{f2}(c) when $J_H \ne 0$. 
		(c) Evaluation of model parameters of Case (ii), with $t=t'=1$, $J_p=0$, and $U=6$. 
	}
	\label{f3}
\end{figure}
\par
The perturbation process is rather complicated as both $J_H$ and $J_p$ hybridize the 
electronic states belonging to the same site when there are two electrons. 
The representative two-electron eigenstates of $\ham_0+\ham_d$ (while taking $t_{ij}=0$) are given in Fig.~\ref{f3}(a). 
In the following, we call the state which has $n_A$ and $n_B$ electrons on site-A and B as ($n_A$+$n_B$)-electron state. 
In Case (i), the low energy manifold was confined to those with \textit{one electron per orbital}, 
but here, since $U'$ differs from $U$ only by $2J_H$, 
the one with the doubly occupied single orbital $\cket{\uparrow_a\downarrow_a}$ on either/both of 
the two sites is also included in the low energy manifold, which we denote $\{\cket{m}\}$. 
Therefore, $\{\cket{m}\}$ consists of all (2+2)-electron states, and the excited states, $\{\cket{l}\}$, 
are the (1+3), (3+1), (0+4), (4+0)-electron states (see Fig.~\ref{f3}(a)). 
Besides taking account of hybridization of states, 
we also need to treat the processes differently from Case (i) in classifying them into two groups; 
$\cket{m}-\cket{l}-\cket{m''}-\cket{l'}-\cket{m'}$ has $\cket{m''}$  
and $\cket{m}-\cket{l}-\cket{l''}-\cket{l'}-\cket{m'}$ has $\cket{l''}$ in the middle (see Fig.~\ref{f3}(b)). 
The Hund's coupling generates several extra paths to the second and third hopping processes, 
as it allows the flipping of spins going in and out of $\cket{m''}$. 
This effect is found to suppress in overall both $J$ and $B$. 
\par
After deriving all the matrix elements between the (2+2)-electron states, we project them 
onto the states with $S=1$ on each site consisting of one electron per orbital via ${\cal P}$ as 
$\tilde\ham_{\rm eff}^{(n)}={\cal P}\ham_{\rm eff}^{(n)}{\cal P}$. 
The second order Heisenberg term is 
\begin{equation}
J_2= 2(t^2+t'^2)/(U+J_H). 
\end{equation}
The fourth order biquadratic terms are evaluated separately for each process~\cite{supplement2} and the dominant 
contribution comes from the twisted ring exchange (T) process and 
the (2b4s) ones, which are given for the case of $J_p=0$ as 
\begin{align}
& B_4^{\rm (T)}
       = t^2 t'^2 \bigg( - \dfrac{12}{a^2e} - \dfrac{4}{ae^2} + \dfrac{4}{a^2b} 
          - \dfrac{12}{a^2f} - \dfrac{4}{af^{2}}   \bigg), \\
& B_4^{\rm (2b4s)}
       = \dfrac{t^4+t'^4}{2} \left(\frac{4}{a^3}  +  \frac{12}{a^2d} + \frac{4}{ad^2} + \frac{2}{a^2b} + \frac{2}{ab^2}
        + \frac{3}{a^2c} +  \frac{1}{ac^2} \right),
\end{align}
with $a=U+J_H$, $b=U-J_H$,  $c=U-3J_H$,  $d=U-5J_H$, $e=U-4J_H$, $f=U-2J_H$. 
(The contribution from $J_p\ne 0$ is not large~\cite{supplement2}). 
Figure \ref{f3}(c) numerically evaluates $J$ and $B$ including all the processes up to fourth order\cite{supplement2}. 
For the Heisenberg terms, the contribution from the fourth order, $J_4$, 
is ferromagnetic and suppresses the antiferromagnetic $J_2$.  
Regarding the biquadratic term, the negative contributions from $B_4^{\rm (T)}$ is suppressed 
by the positive contributions from $B_4^{\rm (2b4s)}$, and resultantly, 
the value of $B$ becomes small by one orders of magnitude, compared to Case (i). 
We finally notice that the $J_H\rightarrow 0$ limit of our results in Case (ii) is not connected to 
$U'\rightarrow U$ of Case (i). 
This is because the formulation of Case (i) is valid at $U'\ll U$ 
and the $U=U'+2J_H \sim U'$ region is properly described only in Case (ii). 
The calculation taking the doubly occupied (2+2)-states as $\{\cket{l}\}$ is discussed in Ref.~\cite{mila}. 
\par
From the comparison of Case (i) and (ii), we find that the values of $J$ do not differ much, 
whereas $B$ takes the value comparable to $J$ only in the former case, 
as $J_H$ is usually much smaller than $U$ and $U'$. 
Therefore, the spin-1 system based on the two-orbital Mott insulator represented by Case (ii) does not afford sufficient degree of biquadratic interaction. 
The idea to overcome this issue is given by Mila and Zhang~\cite{4ptbtn_02}, 
who took account of one extra orbital to each site in Case (ii) that is quasi-degenerate but higher in energy by $\Delta$, 
which is shown in Fig.~\ref{f1} as Case (ii').  
They found that second order processes of particles hoppings to the third orbitals on the neighboring sites 
and coming back will give negative sign to the Heisenberg term, and will cancel out the Heisenberg term at that order on the whole. 
There are closed T-shaped paths consisting of two $t_{\mathrm{ex}}$'s and $t$'s, connecting 
one of the degenerate orbitals and one excited states on each site. 
The fourth order process along this T-shaped path gives rise to additional biquadratic term, 
which becomes the same order as $J$ when $t_{\mathrm{ex}}\sim t'$. 
\par
We finally discuss the possible relevance of process-T and the bulk nematic order. 
For the spin-1 to be relevant in a system based on Case (i), 
we need to have the effective ferromagnetic coupling within each dimer, namely between orbital-1 and 2, and between orbital-3 and 4. 
Placing a strong magnetic field is known to be effective\cite{hikihara}. 
One way to realize such ferromagnetic dimers is to have an indirect hopping between dimerized orbitals 
mediated by the extra orbitals placed off the dimer bonds, 
and the Kanamori--Goodenough rule allows the exchanges to become ferromagnetic~\cite{kanamori-goodenough}. 
This picture is close to the above mentioned protocol by Mila and Zhang~\cite{4ptbtn_02}. 
Another way is to consider a toy model, a uniform square lattice with ferromagnetic nearest neighbor exchange interactions, 
and antiferromagnetic next-nearest neighbor ones in the diagonal direction. 
Indeed, this kind of construction is called $J_1$-$J_2$ square lattice model, and is known to yield a nematic order~\cite{nmtc_j1j2_00}. 
While the origin of such nematic order is attributed to frustration, 
we find that the next-nearest neighbor exchange interactions work together 
with the neighboring ferromagnetic exchanges and form process-T, yielding the effective biquadratic term, 
which shall be a microscopic explanation of what is known so far in numerics. 
\begin{acknowledgments}
We thank Karlo Penc and Fr\'{e}d\'{e}ric Mila for discussions.
This work is supported by Grant-in-Aid for Scientific Research (Nos. 17909321, 17924266, 17895051) from the Ministry of Education, Culture, Sports, Science and Technology of Japan.
\end{acknowledgments}

\clearpage
\clearpage
\pagestyle{empty}
\setlength{\textwidth}{1.15\textwidth}
\begin{figure*}
	\vspace{-20mm}\hspace{-30mm}
	\centering
	\includegraphics[page=1]{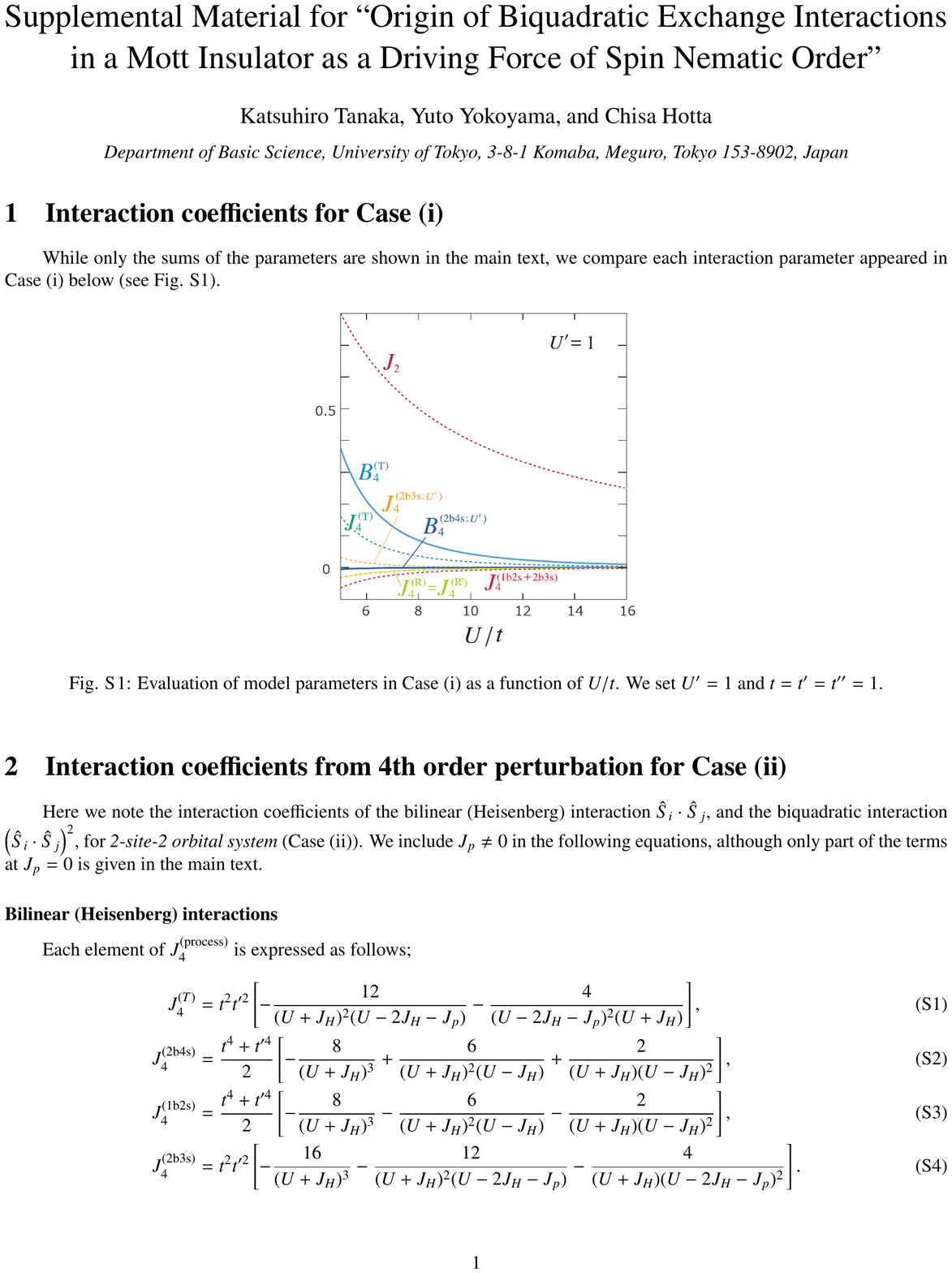}
\end{figure*}
\begin{figure*}
	\vspace{-15mm}\hspace{-30mm}
	\centering
	\includegraphics[page=2]{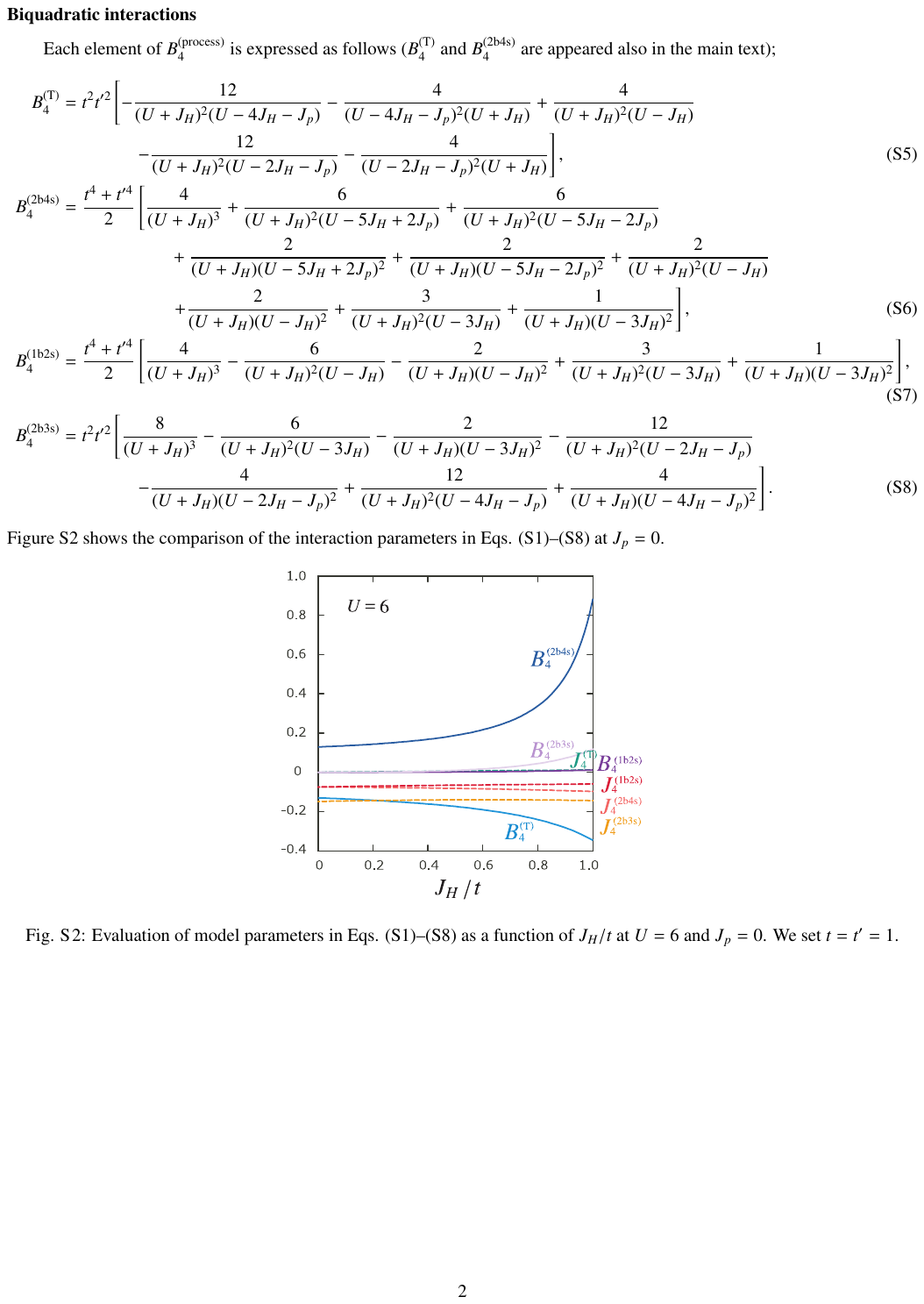}
\end{figure*}
\end{document}